# CCABC: Cyclic Cellular Automata Based Clustering For Energy Conservation in Sensor Networks

Indrajit Banerjee[#], Prasenjit Chanak[*], Hafizur Rahaman[#]

[#]Department of Information Technology
[*]Purabi Das School of Information Technology
Bengal Engineering and Science University, Shibpur, Howrah, India.
[1]ibanerjee@it.becs.ac.in, [2]prasenjit.chanak@gmail.com, [3]rahaman_h@it.becs.ac.in

**Abstract**—*Sensor network has been recognized as the most significant technology for next century. Despites of its potential application, wireless sensor network encounters resource restriction such as low power, reduced bandwidth and specially limited power sources. This work proposes an efficient technique for the conservation of energy in a wireless sensor network (WSN) by forming an effective cluster of the network nodes distributed over a wide range of geographical area. The clustering scheme is developed around a specified class of cellular automata (CA) referred to as the modified cyclic cellular automata (mCCA). It sets a number of nodes in stand-by mode at an instance of time without compromising the area of network coverage and thereby conserves the battery power. The proposed scheme also determines an effective cluster size where the inter-cluster and intra-cluster communication cost is minimum. The simulation results establish that the cyclic cellular automata based clustering for energy conservation in sensor networks (CCABC) is more reliable than the existing schemes where clustering and CA based energy saving technique is used.*

## I. INTRODUCTION

Wireless sensor network (WSN), consisting of thousands of wireless sensors, is bounded due to the limited computational capability, battery power and memory capability of its components. The sensor nodes are deployed in a monitoring area and communicate among themselves following the multi-hop wireless communication. The information received at a node is computed and communicated to the nearest base station [1]. In homogeneous networks, all the sensor nodes are identical in terms of battery energy/power and hardware complexity. The energy saving network design is the major issue in WSN to increase the life time of network nodes.

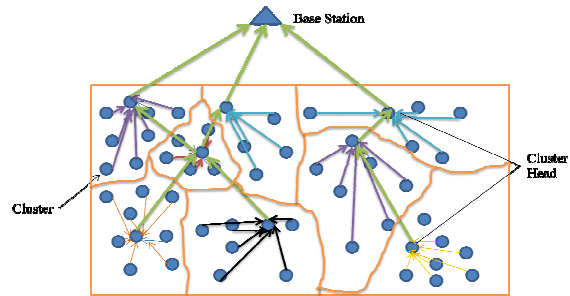

**Fig.1:** Cluster base communication in wireless sensor network

In WSN, a base station (BS) is stationary and the sensor nodes may be movable [2]. The energy loss in a node is very high when a node directly communicates with the base station (Fig.1). On the other hand, in a cluster based node management scheme, a node communicates with the BS through a leader, called cluster head (CH) (Fig. 1). In the clustered scheme proposed so far [2], [3] and [4], each and every node is in active state, therefore, a particular area is monitored by the two or more nodes. The schemes LEACH [2], EEPSC [3], LEACH-C [4], UCCP [5], EECS [6], EEDUC [7] and DDC [8] cannot protect the network nodes from early energy dissipation lead to short span of life. As a node expires within a short time, the new sets of clusters are formed very frequently. This demands massive message exchanges among nodes and, therefore, causes the uncontrolled power dissipation. The cellular automata based techniques [1], [9] and [10] proposed so far recover the common region sensing problem, which ensure that the number of nodes in active state is minimum. The active nodes communicate directly with the base station leading to unwanted energy loss in the network.

In this context, the CCABC technique, proposed in this work, develops a cluster based network management system that ensures full coverage of the sensor network with minimum number of active node. If any active nodes fail to sense, transmit or receive data then this node is declared as a dead node and a neighbouring stand-by node is selected through CCABC scheme for efficient replacement. In the CCABC scheme, the clusters generated are of optimal size, in which the data are aggregated properly for further reduction of overhead in data processing. The member nodes of that optimal cluster size can send their data to the cluster head with minimum energy loss (Fig.2). The proposed scheme also determines the position of the cluster head where each node of the cluster sends their data with minimum energy loss. The CCABC selects a cluster head from the nodes in an energy efficient manner.

The organization of this paper is as follows. Modified cyclic cellular automata (mCCA) are elucidated in Section 2. The mathematical model of the proposed scheme is described in Section 3. In Section 4, we have introduced the proposed algorithms, developed over mCCA. The simulation results are reported in Section 5. Finally we conclude our paper in Section 6.

## II. CYCLIC CELLULAR AUTOMATA

The cyclic cellular automata (CCA) follow a local rule which is same for all states S. Each cell in CCA contents different states from state range S= {0, 1, 2...k-1} [11], [12], [13] and [14]. The integer k is the maximum number of state. In the exiting field $Z^2$ all cells change their states within S.

$$I_t(P) = Z^2 \rightarrow \{0,1,2,...,k-1\} \qquad (1)$$

The CCA generates a spiral structure when cells are changing their states from zero to k-1 as equation number 1. The $I_t(P)$ represents the present state of a cell $P \epsilon Z^2$ at integer time t. If and only if for some given threshold value $\theta$ at *Von Neumann neighbour* set N(x) a cell P changes its state $I_t(P)$ to another state $I_{t+1}(P)$ at time t+1 is shown in equation 2.

$$I_{t+1}(P) = I_t(P) + 1 \bmod k \qquad (2)$$

The threshold value $\theta$ represents the y number of neighboring cell's condition in set N(x) within the field, $y \epsilon N(x)$. The neighboring cells set N(x) are also in k-1 state. The initial state of automaton is said to be primordial soup [11]. A classical model of excitable media was introduced in 1978 by Greenberg and Hastings (GH) [13] and [14] described next.

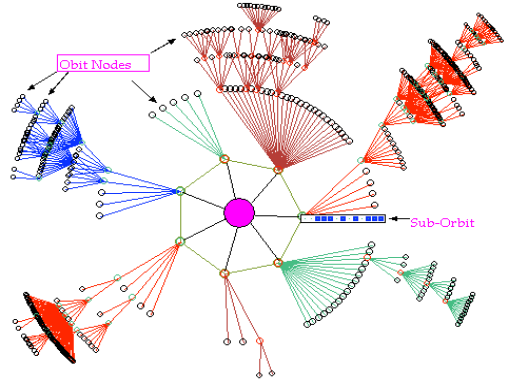

**Fig.2:** Cluster head connected with cluster nodes within a cluster.

**Greenberg-Hastings model (GHM)**

The Greenberg – Hastings model [12] is a simplified cellular automaton that is run in excitable medium. In GHM according to state change rule each cells of the automata changes their state and produced a special type node pattern. The state change rules of the CCA in GHM are described bellow

1. If $\gamma_t(x) = n$, then $\gamma_{t+1}(x = n + 1 \text{ mode } k)$

2. If $\gamma_t(x) = 0$ and at least n neighbours are in state 1 then $\gamma_{t+1}(x) = 1$; otherwise the current state (0) is continued.

Where $\gamma_t(x)$ is cell's condition (or state) at time t and $\gamma_{t+1}(x)$ is the next state of cell at time t+1. In our proposed modified cyclic cellular automata based EERIH [19] we have modified the GSM rules. The EERIH generate special type of nodes pattern and arrange the active nodes into some cluster. This pattern also helps us for routing the data from cluster head to base station in energy efficient manner.

**Proposed modified CCA**

In our modified cyclic cellular automata (mCCA) based scheme every cell changes it state according to the nine neighbours' cells state condition. The state change rules of the cells are defined below [19]:

1. If cell's present state is $\delta_t(p) = n$. where $n > 0 \text{ and } n < I - 1$ , then next state of the cells is $\delta_{t+1}(p) = n + 1$

2. If cell's present state is $\delta_t(p) = n$. where $n = I - 1$, then next state of the cells is $\delta_{t+1}(p) = 0$.

3. If cell's present state is $\delta_t(p) = 0,$ then they check their neighbour cells state and if θ numbers (threshold value) of nodes are present in nonzero state then next state is $\delta_{t+1}(p) = 1$ otherwise they are not changing their state i.e. $\delta_{t+1}(p) = 0$.

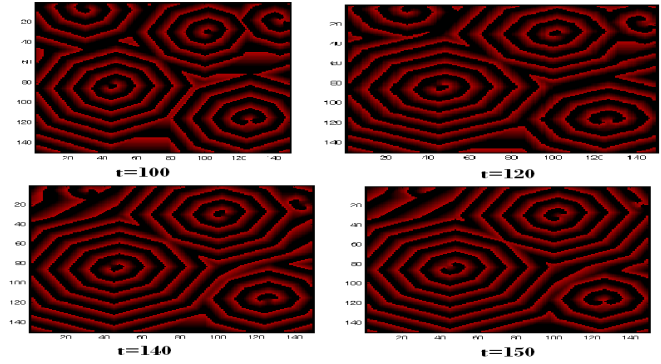

**Fig.3:** Atom ic structure of node pattern generated by CCA

Where $\delta_t(p)$ is the state of the cells at time t and $\delta_{t+1}(p)$ is the state of cells at t+1 time. Number of state is {0, 1…I-1}. With the help of this state change rule we are arranging every sensor node in a pattern of atomic structure. In this scheme every node is changing their state in a time interval and nodes pattern are controlled from the primordial soup. The nodes pattern of the proposed cyclic cellular automata shows in Fig.3 where with time every node is changing their state i.e. spirals are propagated.

## III. MATHEMATICAL ANALYSIS

In this section we have analysed the mathematical model of our proposed scheme energy efficient clustering scheme for wireless sensor networks. Here we calculate effective cluster size where each node sends their data with minimum energy loss. In energy efficient clustering scheme for wireless sensor networks, two types of data communication can take place among nodes in the network. These are intra-cluster and inter-cluster data communication. The inter-cluster data communication is the data transfer between cluster head and base station as shown in Fig. 1. In the proposed scheme the active nodes are self organized into an atomic structure to form clusters (see Section IV). The intra-cluster communication cost is the energy spent by all orbital nodes to send their data to the cluster head. It involves intra-orbit and inter-orbital data communication cost. In intra-orbital data communication, the sub-orbital node transmits their data to the nucleus of the orbit (Fig.2). In inter-orbital data communication, orbits are sending their data to the nucleus of nearest upper layer orbit. If we consider the total sensor network as a single cluster, then inter-cluster communication cost is zero but the intra-cluster communication cost is very high. On the other hand if cluster size is zero, then intra-cluster communication cost is zero but in this case inter-cluster communication cost is very high. With the help of inter-cluster communication cost and intra-cluster communication cost we can formulate an efficient cluster size in cluster based energy efficient wireless sensor networks management scheme, where each node sends their data with minimum energy loss. We have find out the position of the cluster head within a cluster where the communication energy loss is minimal. In wireless sensor network different sensor nodes send their data to base station from different locations. Hence the transmission range of every sensor node is different as well as energy loss of the sensor nodes is also different. In the proposed model every sensor node sends their data with minimum energy loss. These are calculated by energy loss errors in the sensor network. We are also introducing an efficient data aggregation formula of the proposed technique.

**Definition 1:** Let X be the monitoring field, covered by the set of sensor node and $f(\emptyset)$ is the inter-cluster communication cost that is depending on transmission distance and nodes density. Therefore,

$$f(\emptyset) = \int_{e_i \in M_c} \mu \times \beta_s^c((\varepsilon + \omega) + \gamma d_i^n) dd_i \tag{3}$$

The size of data is $\beta_s^c$, in cluster c having s number of nodes. The nodes' density is $\mu$. The $\varepsilon$ is the energy consumed in the transmitter circuit, $\omega$ is the dissipated energy for data aggregation and $\gamma$ is the dissipated energy in the transmitter op-amp. The $e_i$ is the cluster head that collects all data of a single cluster and transmits it to base station. Transmission distance between two cluster head $e_i$ is $d_i$ and n is a path-loss exponent. $M_c$ is the number of member node in a cluster.

**Definition 2:** The intra-cluster communication cost of the node is f(I) that depends on the cluster size and transmission distance between cluster head and cluster member nodes. Therefore,

$$f(I) = \int_{e_i \in P_c} \alpha_i(\varepsilon + \gamma d_j^n) dd_j \tag{4}$$

In above equation $\alpha_i$ is the total bits transmitted along edge $e_i$. $P_c$ is number of nodes in cluster, which are used to collect data from each orbit in a cluster c. The distance between the transmitter and receiver node in a cluster is $d_j$.

**Theorem 1:** Inter-cluster communication cost $f(\emptyset)$ is single-valued and possesses a unique derivation with respect to $\emptyset$ at all points of a WSN region R (in WSN each nodes are connected by multi-hop communication) is called an analytic or a regular function of $\emptyset$ in that region. A point at which an analytic function ceases to possess a derivation is called a singular point of the function.

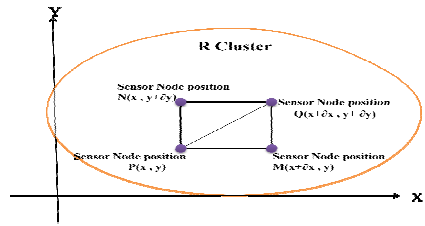

**Fig.4:** Close Region R in WSN, where P and Q are any node position

**Proof-** Let $\omega = f(\emptyset)$ be a single-valued function in a region within the WSN of the variable $\emptyset = x + y$ (Fig. 4). Then the derivative of $\omega = f(\emptyset)$ is defined the inter-cluster communication cost for data transmission between the cluster head,

$$\frac{d\omega}{d\emptyset} = f'(\emptyset) = \underset{\partial\emptyset \to 0}{\text{Lt}} \frac{f(\emptyset + \partial\emptyset) - f(\emptyset)}{\partial\emptyset} \quad (5)$$

provided the limit exists and has the same value for all the different ways in which $\partial\emptyset$ approaches to zero. In sensor network every node is virtually connected to each other, each orbital is virtually connected to upper orbital and cluster heads transmit data to base station with the help of other cluster heads (Fig. 2). Suppose $P(\emptyset)$ is fixed node position within a region R and $Q(\emptyset + \partial\emptyset)$ is a neighbouring node position (Fig. 4). The node Q may approaches towards P along any straight or curved path in the given region R, i.e. $\partial\emptyset$ may tend to zero in any manner and $\frac{d\omega}{d\emptyset}$ to exist.

**Theorem 2:** The inter-cluster communication cost $f(\emptyset)$ is analytic in the cluster region D between two simple close clusters X and X1, then

$$\int_X f(\emptyset)d\emptyset = \int_{X1} f(\emptyset)d\emptyset \tag{6}$$

Where X represents the whole sensor network as a cluster and the X1 is the other cluster within the cluster X.

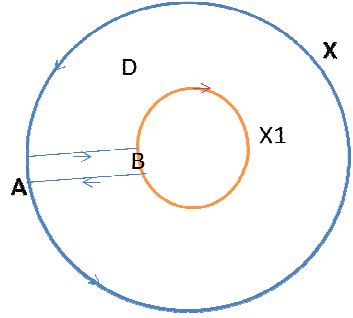

**Fig.5:** X1 is a close cluster under a close cluster X.

**Proof-** We introduce the cross-cut AB in region X. Then $\int f(\emptyset)d\emptyset = 0$. Where the path is as indicated by arrows in Fig. 5; i.e. along AB, X1 in clockwise sense & along BA, X in anti-clockwise sense. Therefore, $\int_{AB} f(\emptyset)d\emptyset + \int_{X1} f(\emptyset)d\emptyset + \int_{BA} f(\emptyset)d\emptyset + \int_X f(\emptyset)d\emptyset = 0$ But, since the integration along AB and BA cancel each other, it follows that $\int_X f(\emptyset)d\emptyset + \int_{X1} f(\emptyset)d\emptyset = 0$. Reversing the direction of the integral around X1 and transposing, we get $\int_X f(\emptyset)d\emptyset = \int_{X1} f(\emptyset)d\emptyset$ here each integration being taken in the anti-clockwise sense. If X1, X2, X3 ...... be any number of close cluster within close cluster X then,

$$\int_X f(\emptyset)d\emptyset = \int_{X1} f(\emptyset)d\emptyset + \int_{X2} f(\emptyset)d\emptyset + \int_{X3} f(\emptyset)d\emptyset \tag{7}$$

*Theorem 2* proves that the total inter-cluster communication cost $f(\emptyset)$ of any number of internal clusters are same as that of whole network.

**Theorem 3:** The inter-cluster communication cost f($\emptyset$) is analytic within a close cluster and if point $a$ is any node position within X, then energy loss at that node is

$$f(a) = \frac{1}{2\pi} \int_X \frac{f(\emptyset)}{\emptyset - a} d\emptyset \tag{8}$$

**Proof-** Let us consider the function $\frac{f(\emptyset)}{(\emptyset - a)}$ which is analytic at all nodes position within X except at $\emptyset = a$ node position with the node position $a$ as centre of cluster and r is the radius of cluster area. We draw a small circle cluster lying entirely within X. Now $f(\emptyset)/(\emptyset - a)$ being analytic in the region enclosed by X and X1. $\int_X \frac{f(\emptyset)}{\emptyset-a} d\emptyset = \int_{X_1} \frac{f(\emptyset)}{\emptyset-a} d\emptyset = \int_{X_1} \frac{f(a+re^\theta)}{re^\theta} re^\theta d\theta = \int_{X_1} f(a + re^\theta) d\theta$ for any nodes on the network, $\emptyset = a + re^\theta$ and $d\emptyset = re^\theta d\theta$. In the limiting form, as the circle cluster X1 shrinks to the node position $a$, as $r \to 0$ we consider every sensor node as a point. The integral approaches to $\int_{X_1} f(a) d\theta = f(a) \int_0^{2\pi} d\theta = 2\pi f(a)$ $f(a) = \frac{1}{2\pi} \int_X \frac{f(\emptyset)}{\emptyset-a} d\emptyset$ In general, $f^n(a) = \frac{n!}{2\pi} \int_X \frac{f(\emptyset)}{(\emptyset-a)^{n+1}} d\emptyset$ X is a large monitoring area, and data is travelling among the nodes in sensor network then $|\emptyset - a| = r$. In this reason $\emptyset$ is unevenly distributed within the close cluster X.

$$|f^n(a)| \le \frac{Mn!}{r^n} \qquad (9)$$

Where M is the maximum value of $|f(\emptyset)|$ on cluster X.

In-order to find out the actual position of the cluster head in the network, we have divided the whole network into equal size of clusters. The suitable position of the cluster head can be determined according to the following theorem.

**Theorem 4:** The analytic function inter-cluster communication cost $f(\emptyset)$ within the close cluster region is average at the centre position of a cluster region. In this region f(I) is very small. The centre position, where inter-cluster communication cost $f(\emptyset)$ is average and intra-cluster communication cost f(I) is small, is the cluster head location.

**Proof**- When we consider the cluster of sensor node as a circle then inter-cluster communication cost of the cluster node $f(\emptyset)$ is average in the centre point of the cluster. Because inter-cluster communication cost $f(\emptyset)$ depends mainly on the distance between cluster head and base station. On the other hand intra-cluster communication cost $f(I)$ also depends on the member nodes to cluster head distance. If we select cluster head as a nearest node of the base station then the inter-cluster communication cost is minimum but within the cluster member nodes and cluster head, distance is increased (so the value of f(I) is high), therefore, total energy loss by the cluster $f(\emptyset) + f(I)$ is increased. Within cluster long distance cluster member nodes lose more energy. So we are going to select cluster head's position at the median of the cluster.

Theorem 1 describes inter-cluster communication cost f(∅), which is an analytic function, because this function follows the necessary and sufficient condition of an analytic function. We are selecting some nodes position of whole network according to Theorem 2 and start the CCA spiral propagation for the selection of nodes' position. With time the CCA spiral propagation cover the entire network and therefore f(∅) value is increased and f(I) value is decreased. When the inter-cluster communication cost and intra-cluster communication cost is equal i.e.f(∅) = f(I), the spiral propagation (see Section IV) will stop at that point. This is the optimal size of the cluster where data are aggregated properly according to the equation no. 11. The member nodes of that optimal cluster size can send their data with minimum energy loss. With the help of Theorem 3 we are going to calculate inter-cluster communication cost in each sensor node and determine the position of the nodes from where CCA spiral start to propagate. Theorem 4 determines the position of the cluster head where each node of the cluster sends their data with minimum energy loss.

**Data Aggregation Model**

Data aggregation on a sensor network depends on data correlation. With the help of data correlation we can aggregate efficient amount of data from the nodes. The aggregated data is then transmitted to cluster head (CH). A large amount of energy is wasting due to transmission of same type of information to the base station. When density of the active nodes increases to provide fault tolerant feature in the sensor network [16], large number of number of nodes cover a particular area which sense similar information. If density of the nodes is very high, then a large amount of energy is wasted for transformation of same information through multi-hop sensor network. Different types of approaches have been proposed to model the correlation of data; one of them, entropy-base model is very popular. The entropy-base data correlation and compression algorithm is described in [17].

$$B_s(d_0) = b_0 + (s-1)(1 - \frac{1}{\frac{d_0}{c}+1})b_0 \qquad (10)$$

Here $d_o$ is the inter-node distance and $b_o$ is the number of bits generated by each source, the constant parameter c characterizes the spatial data correlation. $B_s(d_0)$ is the number of compressed bit messages generated by the cluster head in a s-node cluster. The entropy-base model is applied in the CCABC scheme, as the model aggregates data accurately and efficiently. In CCABC, we have modified the equation no. 18 in-order to decrease the data aggregation error rate.

$$B_s(d_0) = b_0 + (s-1)(1 - \ln 2 e^{-\frac{d_0}{\sigma+s}})b_0 \tag{11}$$

$\sigma$ represents the minimum size of the cluster. Other parameters have the same meaning as in equation no 18. The correlation error reduces in our proposed equation. If the cluster size increases the data correlation error increases. So we consider here the optimal cluster size.

## IV. CCABC ALGORITHM

The modified CCA (see Section II) is used to create clusters in sensor network. During the change of state the spirals are decomposed randomly in a time interval and generate wave pattern. In this way nodes are self-organized into an atomic structure (AS). In an atomic structure the nodes are distributed into orbits and nucleus. The orbit is also divided into sub-orbits. The proposed cluster formation algorithm is as follows:

**CCABC (Algorithm 1): Cluster Generation**

    Insert all nodes into S (array of nodes)
    CS (Cluster Set) = Null (empty set)
    **WHILE** S! = Null **DO**
        Calculate Nodes energy from Algorithm 3
        Set CC = Si (single Cluster)
        Set Inter cluster Communication Cost f ($\emptyset$) = 0
        Calculate Intra Cluster Communication Cost f (I)
        Orbital to orbital distance from Algorithm 2 when    mCCA is starting
        Take some nodes position Pi where mCCA starts to spiral propagation within Si
        Check f($\emptyset$) and f (I) after some specific time slot
        **IF** f ($\emptyset$) = f (I) **THEN**
            Stop spiral propagation
            Insert into CS (cluster set)

      **ELSE**

          Carry on spiral propagation

      **END IF**

    Start data aggregation and data transmission

**END WHILE**

**CCABC (Algorithm 2): Orbital to Orbital distance Calculation**

Set $r_{ca}$ is current transmission range

Set $r_{ca} = r_{max}$

Finding neighbour ($r_{ca}$, i, j)

Calculate D (Density)

Calculate $r_{min}$

Calculate $r_{od}^2$

Set the orbital distant $r_{od}^2$

**CCABC (Algorithm 3): Verification Algorithm**

**IF** nodes Energy is less than threshold **THEN**

    Nodes is dead

    **IF** Ti = 0 **THEN**

        CCA Is rotated

    **ELSE**

        Decrement time

    **END IF**

**END IF**

    In proposed scheme all nodes have two states; the active state and the stand-by state. In active state a node senses the data and transmits it to cluster head. The stand-by nodes are in sleeping mode. Any one node from the nucleus is acting as cluster head. Each sub-orbit transmits their data to the upper sub-orbit within an orbit (Fig. 2). In each orbit, the sub-orbit, which is nearest to the nucleus, collects other sub-orbits data. Then the data is aggregated and transmitted to next orbital. Nearest orbit of nucleus transmits data to the cluster head in nucleus. Cluster head collects data and aggregates and transmits it to the base

station. The position of the base station is fixed. Orbital nodes are changing their states randomly and repeatedly.

The whole network energy as well as the energy utilization of the nodes is divided into two parts; one is data sensing and another is data transmission. In initial condition the network is divided into some clusters of actual size depending on intra-cluster communication cost f(I) and inter-cluster communication cost $f(\emptyset)$ as discussed in *Theorem 2*. The cluster nodes are arranged like atomic structure with the help of modified CCA, cluster head (CH) is selected from nucleus. At the nucleus few nodes are in active state, one out of them will act as a CH and rest will behave as normal node. At the nucleon the f(I) and $f(\emptyset)$ is minimum at this point according to *Theorem 4*. These nodes are also changing their state from active to stand-by. When CH changes its state or its energy reaches to a threshold value, any other active node from the nucleus may be designated as a CH for next round.

The transition phase is divided into two parts; one is data collection part and another is data transmission part. In the data collection part active nodes are sensing data from their monitoring area and in transmission phase selected sub-orbital nodes and cluster heads are collecting other cluster member data, aggregating and transmitting the same. In nucleus, one of the nucleon nodes is selected as a cluster head by CCABC. A sufficient amount of area is covered by the cluster head. Inter-cluster communication costs also minimum in CCABC scheme as shown in *Theorem 2*.

In CCABC, data bits are aggregated in different orbits. In CCABC model the data are propagating form lower orbit to the upper orbit. Finally cluster head gets the aggregated data from nearest orbit. Hence cluster head's data aggregation and receiving loads are distributed in different orbital node, therefore, receiving energy loss in CCABC cluster head is less. The transmitting, receiving energy losses calculating formula and CCABC orbital to orbital distance calculation formula are defined below.

## V. SIMULATION RESULT

The whole simulations are done in MATLAB. The whole simulations are divided into two parts, one is the cluster formation through spiral propagation and another is data transmission with energy calculation phase. In data transmission and energy calculation phase every node checks their present energy and compares with its threshold value. If present energy is less than the threshold, then the node is declared as a dead node. The cluster formation is based on internal knowledge of each node and massage passing is not required. The spiral propagation will be stopped when the clusters are formed where the inter-cluster and intra-cluster communication cost is minimum. The nodes are sensing data, aggregating and transmitting it to the cluster head. In each round, cluster member nodes transmit their sensed data for five times to the cluster head. The CH collects this data and then it is transmitted after aggregation to the base station. After data transmission round each node calculates their remaining energy according to transmission and receiving energy loss equations which is describe in [19]. These two processes are running continuously for the entire life span of the network.

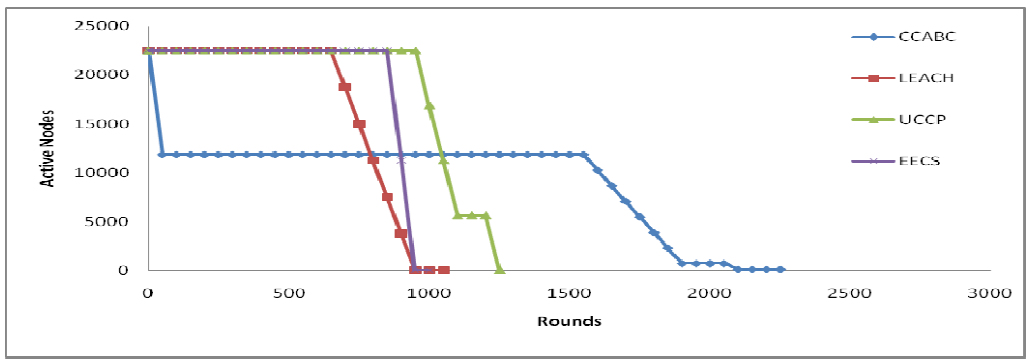

**Fig. 6:** Number of Active Nodes in CCABC

We have taken 150×150 matrix and number of state k=15. In this matrix we have applied mCCA. The CCABC nodes are self-organized. The Fig. 6 shows the total number of active nodes per round in a network. In the CCABC, sensor node survives longer time than the other automata based algorithm and cluster based algorithms e.g. LEACH, UCCP, M-LEACH and EECS. In this CCABC, clusters are generated without message transmitting and cluster heads are selected on the basis of local information.

Whereas, the other popular clustering technique like LEACH, UCCP, EECCP some amount of energy are spent by the message passing for cluster generation. In CCABC, message over heading problem can be recovered, which is present in LEACH clustering technique. Table 1 show the parameter values which is used in simulation.

**Table 1**
**Simulation parameters**

| | |
|---|---|
| Sensor Deployment Area | 150×150 |
| Base Station Location | (50,175) |
| Number of node | 22500 |
| Data Packet Size | 800bit |
| Initial Energy | 0.5J |
| Stand-by state energy loss | 0.00006J |
| Energy per bit spent by the transmitter circuit ($e_t$) | 50 nJ/bit |
| Amplifier energy ($e_d$) | 10 pJ/bit/m$^2$ |

In CCABC simulation technique, initial energy of a node is 0.5J and packet size of each message is 100 bytes is fixed in whole simulation. The base station position is fixed. In each round orbital nodes collect data and send message with the help of TDMA (Time division multiple-access) MAC protocol [18]. In each round, nodes are sending five data message to the cluster head. Cluster head collect their data and send data message to the base station. For data aggregation, energy spent is 5nJ/bit/message [2].

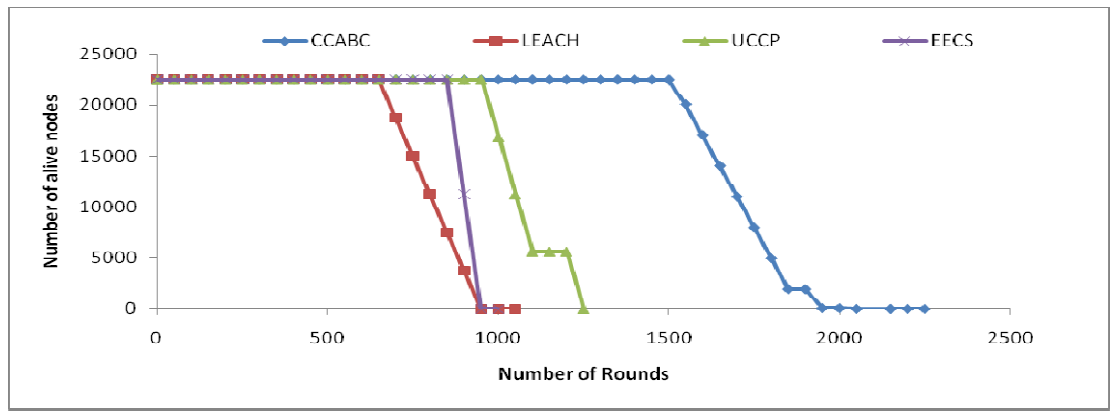

**Fig.7**: Life time of the nodes in CCABC

When the clusters are generated by CCABC scheme at the beginning, the active nodes are 10652 in number. These active nodes cover whole network field, and number of stand-by nodes are 11848. We compare CCABC with other clustering algorithm LEACH, UCCP (Unified Clustering and Communication Protocol [5]), EECS (Energy Efficient Clustering Scheme [6], [14]. The result shows that (Fig. 7.) in CCABC, death of the first node occurs after 1560 round which is better than other existing algorithm. The CCABC extend network life time 41% over UCCP, 64% over LEACH and 54.55% over EECS.

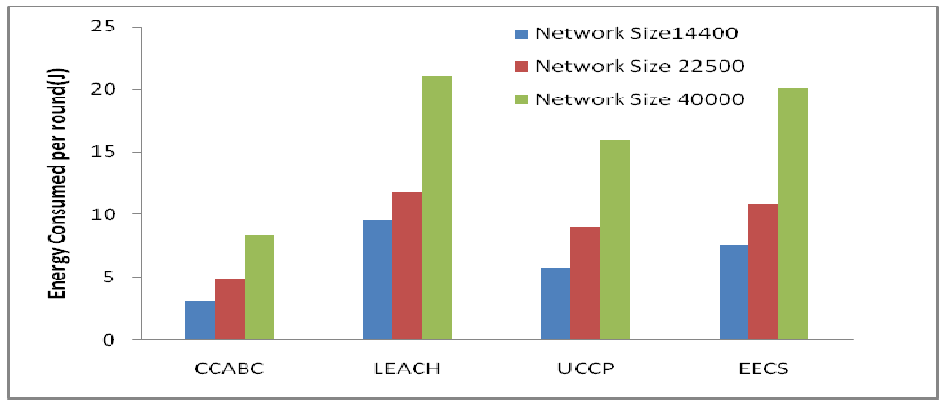

**Fig.8:** Average energy consumed per round

The Fig. 8 represents the average energy consumption per round for three different network sizes. These statistics are collected using 1500 independent rounds with no dead nodes in the network. It can be observed that CCABC outperforms over all other protocol because it generates clustering with the help of cyclic cellular automata and a large number of nodes are in stand-by state. On the other hand CCABC reduces message overheads compared to other technique.

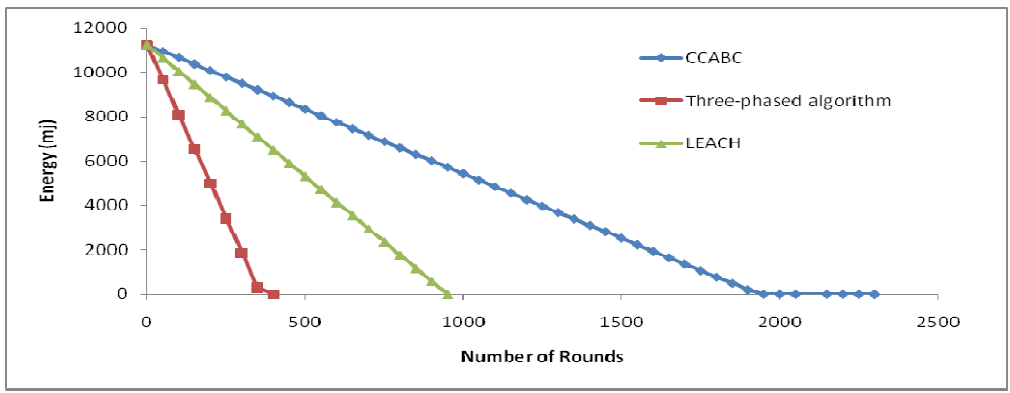

**Fig. 9:** Total amount of energy of the network in energy CCABC

Total energy spent by the sensor network in CCABC, three-phased algorithm and LEACH are compared in Fig. 9. The CCABC saves 60.5305(J) energy per round compared to three-phased algorithm and 13.898(J) energy per round compared to LEACH. The better energy utilization of CCABC scheme increases network's lifetime 81.83 %  more compared to three-phased algorithm and 56% more compared to LEACH.

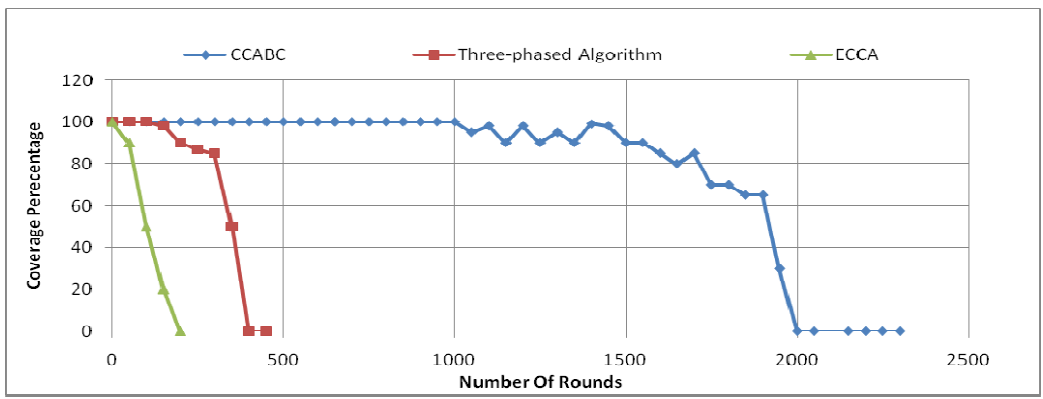

**Fig. 10:** The network coverage in CCABC vs. Three-phased algorithm and ECCA

The area which is monitored by the sensing range of all active nodes is called coverage area. In the Fig. 10 we have compared percentage of coverage between CCABC and three-phased algorithm. We have got better result compared to three-phased algorithm. The experimental results confirm that the network coverage is approximately 40% in three-phased algorithm and 20% in ECCA [9] algorithm, whereas, the CCABC achieves up to 80% of network coverage.

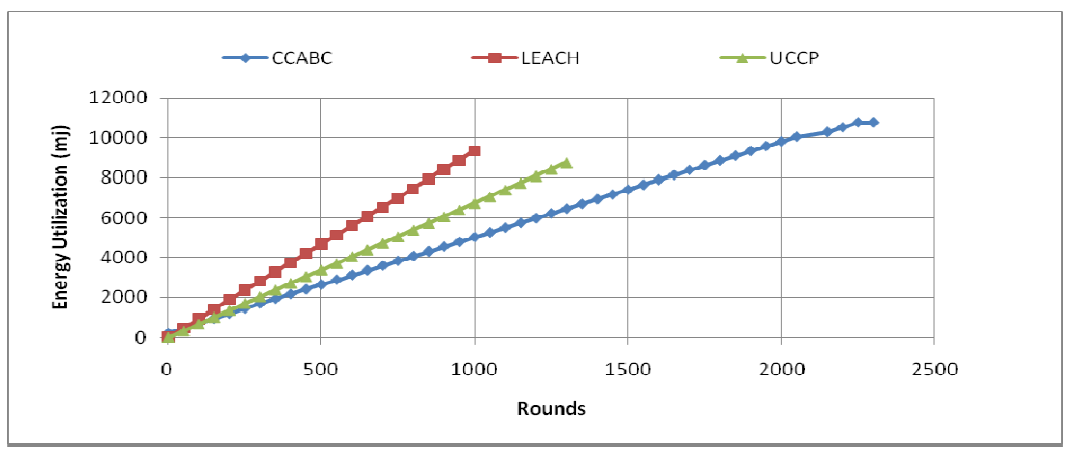

**Fig. 11:** Energy utilization of the sensor network

In CCABC, sensor nodes send their data with minimum energy loss compared to other existing algorithm LEACH, UCCP. In CCABC 52.82% of nodes are in stand-by state, loses minimum amount of energy. 47.34% nodes are in active state. The energy utilization for topological management in different network is shown in Fig 11. The energy utilization for topological management in LEACH and UCCP is 83% and 77.9% respectively. Whereas, in cluster based energy efficient wireless sensor networks management scheme 60.9% energy uses for topological management.

## VI. CONCLUSION

In this paper we have designed an energy efficient sensor network with modified cyclic cellular automata based clustering. Modified cyclic cellular automata (mCCA) splits the whole network into some effective size clusters. The mCCA also ensure maximum coverage in the network with minimum active nodes. In CCABC we have determined an optimal cluster size where the node sends their data with minimum energy loss. An efficient cluster head position is also determined in CCABC. Here we have proposed a new effective data aggregation model which is used by the cluster heads before data propagation. The simulation results established that the proposed scheme is better compared to other popular clustering algorithms. The mobile network model works with CCABC, may show better performance in real application. The CCABC is suited for VLSI implementation and, therefore, the proposed management scheme can be implemented with low cost hardware.

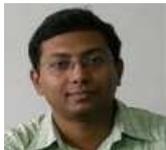
**Indrajit Banerjee** is an assistant professor in the Information Technology Department at Bengal Engineering and Science University, Shibpur, India. He got the bachelor degree in mechanical engineering from Institute of Engineers, India. He received his masters in Information Technology from Bengal Engineering and Science University in 2004. He is currently pursuing his Ph. D. in Information Technology in Bengal Engineering & Science University. His main research interests are cellular automata, wireless ad hoc and sensor network.

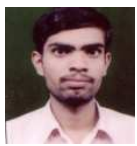
**Prasenjit Chanak** received his B.Tech degree in Information Technology from Institute of Engineering and Technology, U.P., India in 2007. He received his masters degree in Information Technology from Bengal Engineering and Science University in 2011. His main research interest is wireless ad hoc sensor network.

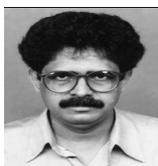
**Hafizur Rahaman** received the B.E. degree in electrical engineering from Bengal Engineering College, Calcutta University, Calcutta, India, in 1982 and the M.E. degree in electrical engineering and Ph.D. degree in computer science and engineering from Jadavpur University, Calcutta, in 1988 and 2003, respectively. During 2006–2007, he visited the Department of Computer Science, Bristol University, Bristol, U.K., as Postdoctoral Research Fellow. He is currently chairing the Department of Information Technology, Bengal Engineering and Science University, Howrah, India. His research interests include logic synthesis and testing of VLSI circuits, fault-tolerant computing, Galois-field-based arithmetic circuits, and quantum computing. Dr. Rahaman has served on the Organizing Committees of the International Conference on VLSI Design in 2000 and 2005 and as the Registration Chair for the 2005 Asian Test Symposium, which was held in Calcutta.